\documentclass[9.96pt,conference]{IEEEtran}
\IEEEoverridecommandlockouts
\usepackage{cite}
\usepackage{amsmath,amssymb,amsfonts}
\usepackage{algorithmic}
\usepackage{graphicx}
\usepackage{textcomp}
\usepackage{xcolor}
\usepackage[caption=false,font=normalsize,labelfont={scriptsize},textfont=sf]{subfig}
\usepackage[detect-all]{siunitx}
\usepackage{acronym}\usepackage[acronym,style=long3col,nomain, toc]{glossaries}
\usepackage{listings}
\usepackage[hidelinks]{hyperref}
\usepackage{orcidlink}
\usepackage[none]{hyphenat}
\usepackage[left=0.673in, right=0.727in, top=0.760in, bottom=1.103in]{geometry}
\def\BibTeX{{\rm B\kern-.05em{\sc i\kern-.025em b}\kern-.08em
    T\kern-.1667em\lower.7ex\hbox{E}\kern-.125emX}}
\begin{document}

%
\newacronym[]{eh}{EH}{Energy Harvesting}
\newacronym[]{iot}{IoT}{Internet of Things}
\newacronym[]{ml}{ML}{Machine Learning}
\newacronym[]{fn}{FN}{Forest New}
\newacronym[]{gn}{GN}{Grassland New}
\newacronym[]{go}{GO}{Grassland Old}
\newacronym[]{lpwan}{LPWAN}{Low Power Wide Area Network}
\newacronym[]{wdt}{WDT}{Watchdog Timer}
\newacronym[]{sim}{SIM}{Sensor Interface Manager}
\newacronym[]{rtc}{RTC}{Real-Time Clock}
\newacronym[]{alp}{ALP}{Application Layer Programming}
\newacronym[]{d7a}{D7A}{DASH7 Alliance Protocol}{}
\newacronym[]{d7advp}{D7AAdvP}{D7A Advertising Protocol}
\newacronym[]{d7actp}{D7AActP}{D7A Action Protocols}
\newacronym[]{wsn}{WSN}{Wireless Sensor Network}
\newacronym[]{teg}{TEG}{Thermoelectric Generator}
\newacronym[]{prr}{PRR}{Packet Reception Ratio}

\title{Experiences with Sub-Arctic Sensor Network Deployment\\
}

\author{
    \IEEEauthorblockN{Priyesh Pappinisseri Puluckul\orcidlink{0000-0003-4145-9443}, Maarten Weyn\orcidlink{0000-0003-1152-6617}}
    \IEEEauthorblockA{\textit{Department EICT, IDLab-imec,}\\ \textit{University of Antwerp, 2000, Belgium}\\
    \{priyesh.pappinisseripuluckul, maarten.weyn\}@uantwerpen.be}
}

\maketitle

\begin{abstract}
This paper discusses the experiences gained from designing, deploying, and maintaining low-power Wireless Sensor Networks (WSN) in three geothermally active remote locations in Iceland. The network was deployed for environmental monitoring and real-time data collection to assist in investigating the impact of global warming on the (sub)Arctic climate and the resulting carbon release from the region. Functional networks with more than \num{50}  sensor nodes from three sites with extreme weather conditions and hard-to-access terrain have been collecting data since 2021. The networks employ primary cell-powered wireless sensor nodes equipped with \acrfull{d7a} for low-power data transmission and solar-powered \acrshort{d7a}-cellular gateways for the backend connection. The WSNs have so far achieved over three years of uptime with minimal maintenance required throughout this period. We present a detailed discussion of different network components, their architecture, and the networks' overall performance and reliability.
\end{abstract}


\section{Introduction}
The Arctic ecosystem is a large reservoir of carbon, developed over millions of years~\cite{arctic_ecosystem}. Climate changes and global warming cause the ecosystem to release a considerable amount of stored carbon, significantly impacting the carbon cycle. It is estimated that close to \num{30}\% of the global carbon is stored in northern high latitude regions, despite it covering only \num{5}\% of the global soil surface~\cite{scharlemann2014global}. Therefore, understanding the future release of carbon from the Arctic system and its impact on climate is crucial. However, estimating the effect of future warming requires soil temperature manipulating experiments that are often costly and not scalable~\cite{bjarni_forhot}. In this regard, the ForHot site located in the village of Hverager$\eth$i in southern Iceland offers a geothermally warmed soil~\cite{forhot}. This organic warming due to geothermal activity can be used as a large-scale controlled environment to study the impact of global warming and future carbon release, thereby providing a natural laboratory. Nevertheless, continuous measurement of different environmental parameters is essential to understand how the ecosystem responds to the warming.

\glspl{wsn}  have been proven to be an effective tool for supporting long-term environmental monitoring applications, particularly in situations where frequent visits for data collection are challenging. Many successful deployments of \glspl{wsn} for environmental monitoring have been reported in the past~\cite{wsn1, wsn2}. To assist the investigation of climate change and its impact on the Arctic ecosystem, we designed and deployed  \glspl{wsn} in ForHot, which provided real-time access to numerous physical parameters from three remote locations. By utilizing primary cell-powered sensor nodes with ultra-low energy consumption and solar-powered gateways, the networks have achieved over three years of lifetime. In this paper, we outline the experiences with these network deployments, along with the details on the long-term reliability and performance of the network.

\begin{figure}[t]
\centering
    \subfloat[]{%
    \includegraphics[width=1.5in]{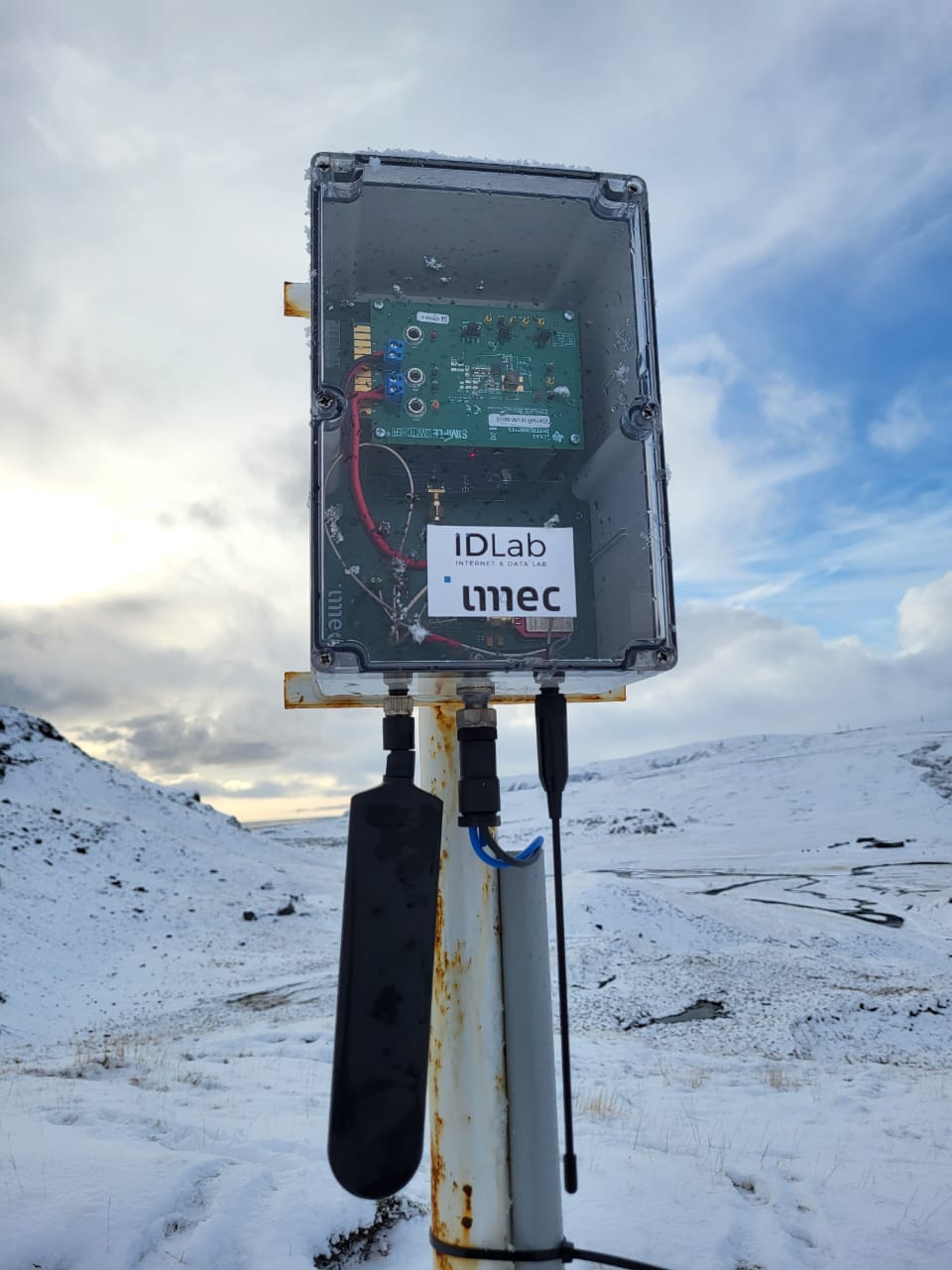}
    \label{fig_gateway_in_winter}}
    \quad
    \subfloat[]{%
    \includegraphics[width=1.5in]{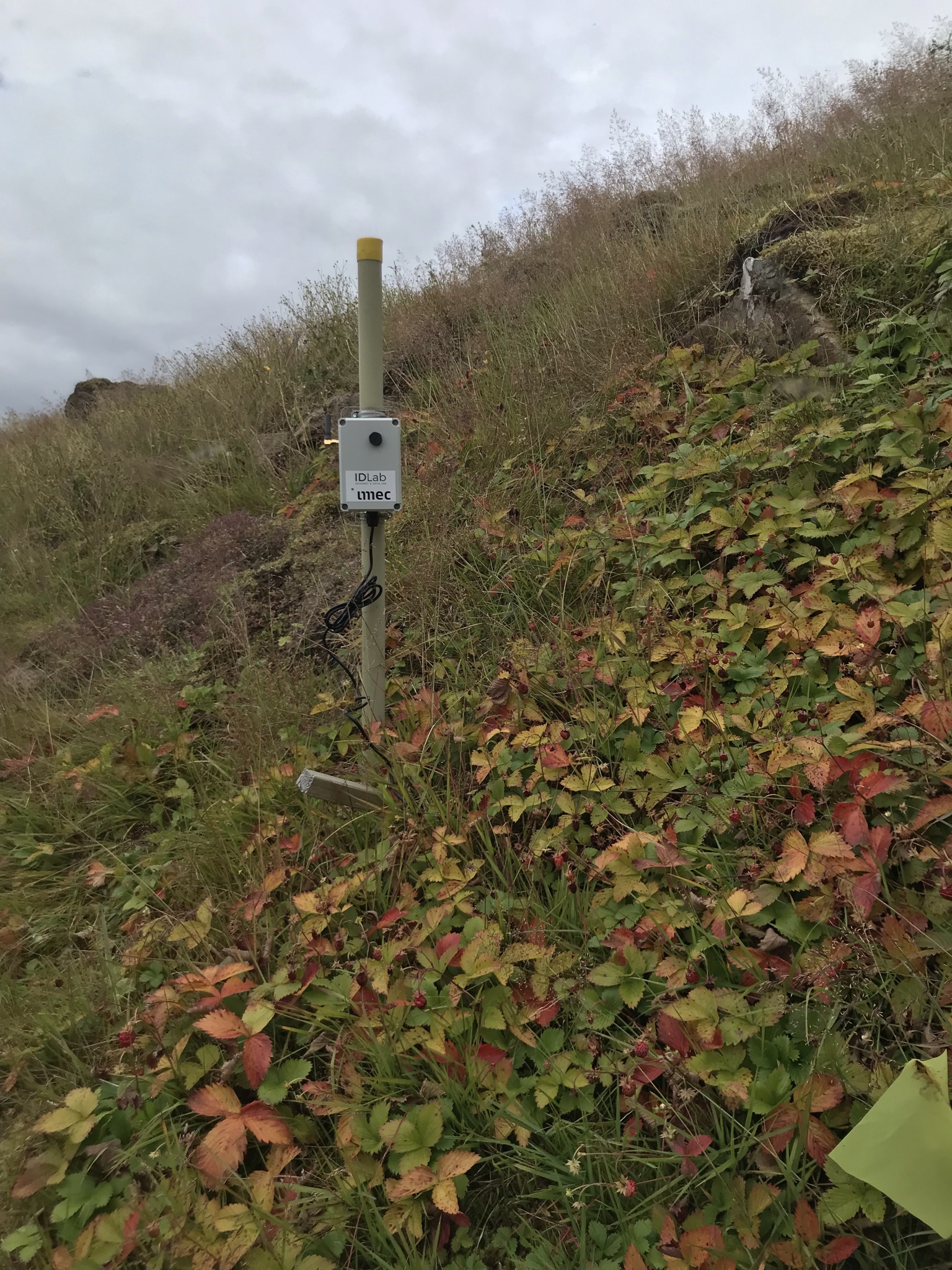}
    \label{fig_sensor_node_deployment}}
    \caption{Deployments at the geothermally warmed locations within the ForHot site in Iceland.}
    \label{fig_introduction_figure}
\end{figure}

The remainder of this paper is organized as follows: Section~\ref{section:deployment_site} provides an overview of the three deployment sites. Section~\ref{sec:network_components_and_design} presents a detailed discussion of different network components and their design. Section~\ref{section:deployment} discusses the deployment activities and timeline. Section~\ref{section:discussions} includes an elaborated discussion on the outcome of the deployment, including data collection, network reliability, battery lifetime, and the lessons learned. Finally,  Section~\ref{section:conclusion} presents the conclusions of the research.

\section{Deployment Sites}\label{section:deployment_site}
\begin{figure}[h]
    \centering
    \includegraphics[width=2.5in]{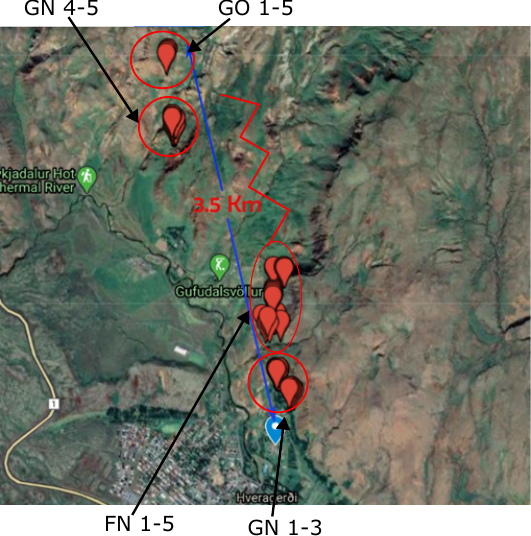}
    \caption{Google Earth map showing the three sites in Iceland.}
    \label{fig:sites_google_earth_view}
\end{figure}

The ForHot site is divided into different sections based on the nature of warming as \gls{fn}, \gls{gn} and \gls{go}~(Fig.~\ref{fig:sites_google_earth_view})~\cite{bjarni_forhot}. The soil at the \gls{go} site has been warmed since 1963. Whereas, the warming at \gls{fn} and \gls{gn} started only after a massive earthquake in 2008, which affected the nearby geothermal systems and consequently started the warming process~\cite{bjarni_forhot}. Only the \gls{gn} and \gls{go} sites are considered for the deployment and \gls{fn} is excluded.  The \gls{gn} site is further divided into two as \gls{gn}~1-3 and \gls{gn}~4-5. These sites are independent but are collectively named \gls{gn} and are recently warmed. The \gls{gn} 4-5 and \gls{go} sites are characterised by the presence of thick and tall grass, whereas there are spruce trees present at the entrance of the \gls{gn}~1-3. Additionally, all the sites have tough terrains and steep hills spread over a wide area. The sites receive heavy snowfall during the winter and can be under snow cover for most of the winter months. The distance between \gls{gn}~1-3 and \gls{go} is around \SI{3.5}{\kilo\meter} but there are no direct connecting paths between the two. The \gls{gn}~4-5 falls around in the middle of both the sites. It may be noted that while the Google map shows the sites close to each other, they are obstructed by hills.  

Each site is  divided into five plots~(\gls{gn}/\gls{go}~1-5) and each plot is further divided into six transects~(A-F) of \SI{50}{\centi\meter}~x~\SI{50}{\centi\meter} size. These transects are created based on the warming level of the soil; transect $A$ is unwarmed. Taking transect $A$ as a reference,  $B$ is warmed approximately by \SI{1}{\celsius}, $C$ by \SI{3}{\celsius}, $D$ by \SI{5}{\celsius} and $F$ by \SI{10}{\celsius}. It is important to note that these warming levels are based on measurement campaigns conducted between 2013 and 2016, with soil temperatures sampled at a depth of \SI{10}{\centi\meter}~\cite{bjarni_forhot}.

For the ecosystem research, we were required to measure soil temperature at \SI{10}{\centi\meter} from   \num{54} locations (3x6 in \gls{gn}~1-3, 2x6 in \gls{gn}~4-5 and 4x6 in \gls{go}~1-4), weather condition at \gls{gn}~4-5 and water content at each site. 

\section{Network Components and Design}
\label{sec:network_components_and_design}
This section gives an overview of the architecture of the overall network and the hardware-software design of different network components.
\subsection{Choices of Wireless Communication}
We considered three possible \gls{lpwan} standards at the initial stage of design- LoRaWAN, NB-IoT and \gls{d7a}. LoRaWAN can provide long-range connectivity at low data rates with extremely low power consumption. NB-IoT, on the other hand, co-exists with cellular networks and can enable a wide range of connectivity. However, NB-IoT has a relatively higher power consumption than LoRaWAN devices~\cite{nbiot-lora}. \gls{d7a} is another communication protocol for low-power wireless sensors and actuators in the ISM band. \gls{d7a} can enable connectivity over \SI{2}{\kilo\meter} in outdoor conditions. Therefore, \gls{d7a} is not truly an \gls{lpwan} technology, but a mid-range protocol.  \gls{d7a} has comparable power consumption levels to LoRa and can provide low-power downlink through an intermittent wake-up protocol~\cite{weyn2015dash7}.

At the time of planning the FutureArctic deployment, no NB-IoT network was available in Iceland. As a result, NB-IoT was excluded from further consideration. Out of the remaining two, LoRaWAN has an advantage of long-range over \gls{d7a}. Whereas, \gls{d7a} has the advantage of low-power wake-ups, which is important for establishing downlink communication to the devices.  \gls{d7a} requires no periodic synchronization to enable downlink transmission.   During the initial planning, it was decided that the network would also provide real-time ground-truth soil temperature data for drone-based remote sensing via on-demand querying of the sensor nodes. Therefore, the low-power bidirectional capability of \gls{d7a} was considered an advantage, which led to the selection of \gls{d7a} over LoRa.

\subsection{Wireless Sensor Nodes}
\begin{figure}[t]
\centering
    \subfloat[]{%
    \includegraphics[width=1.6in]{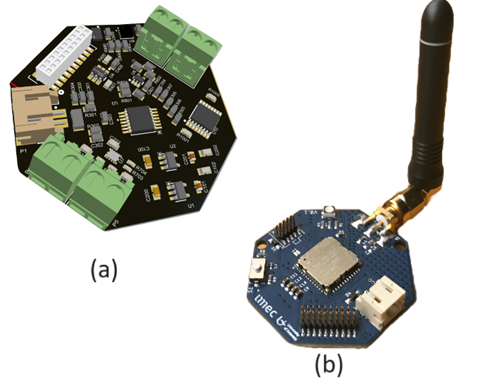}
    \label{fig:sensor_node_components}}
    \quad
    \hspace{-2em}
    \subfloat[]{%
    \includegraphics[width=1.6in]{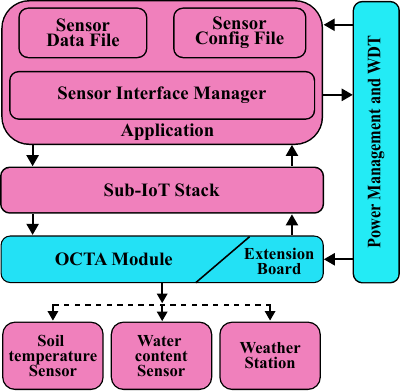}
    \label{fig:sensor_node_architecture}}
    \caption{A picture of the \protect\subref{fig:sensor_node_components} hardware modules: (a)~extension board  (b)~OCTA board and \protect\subref{fig:sensor_node_architecture} a block diagram of the hardware stack.}
\end{figure}
The requirement for long-term unattended operation makes the power consumption of the network components the most critical design consideration. Therefore, hardware-software architectures with the lowest possible energy consumption were employed, while providing an interface to a soil temperature sensor~(DS18B20: 1-Wire), water content sensor~(Meter Group TEROS 10: SDI-12), and weather station~(Meter Group ATMOS 41: SDI-12). Moreover, it must be reliable, modular, and easily re-configurable to provide a seamless sensor interface so that anyone without much technical know-how can deploy the devices. Modularity is therefore another important aspect of the design, especially given that the network will be deployed miles away, making troubleshooting and maintenance visits costly. 

The hardware architecture of the sensor nodes is based on the low-power OCTA board\footnote{https://shorturl.at/WQvJE}. OCTA uses a CMWX1ZZABZ-091 radio modem from Murata, which can be programmed as a \gls{d7a} transceiver.  GPIOs and peripheral connections are available through a 10-pin connector on the OCTA board which can be used to connect external peripherals or extension boards~(Fig.~\ref{fig:sensor_node_components}.b). We use an extension board to provide additional sensors and interfaces~(Fig.~\ref{fig:sensor_node_components}.a). The extension board houses a temperature-humidity sensor, a watchdog timer, flash memory for data buffering, a configurable power supply ($V_{OUT}$ = \SI{5}{\volt}, \SI{3.3}{\volt}), an ultra-low power ADC. The configurable power supply allows the device to accommodate external sensors with different voltage requirements. The entire hardware of the node is powered using a \SI{19}{\ampere\hour} $LiSOCl_{2}$ primary cell~\cite{liscol2_cell}. A picture of the deployed sensor node is shown in Fig.~\ref{fig_sensor_node_deployment}.

The sensor nodes run the open-source \gls{d7a} implementation from SubIoT\footnote{https://github.com/Sub-IoT} to enable \gls{d7a} connectivity along with additional application firmware to support and manage sensors and remote access.  The application layer of the node is required to perform three primary tasks (i) sample sensors at predefined intervals and transmit the data (ii) provide a seamless interface to sensors and (iii) manage downlink messages and respond according to the specific downlink requests. The hardware and software stack of the sensor node is depicted in Fig.~\ref{fig:sensor_node_architecture}. Since \gls{d7a} is a file-based communication protocol, all the sensors and peripherals on the node can be accessed over-the-air using standard file operations~\cite{weyn2015dash7}. Therefore, at the application layer, we define a file for sensor data, allowing access to current measurements, and a configuration file to update parameters like sensor type, sampling rate, and resolution. The sensor type parameter in the configuration file allows changing the sensor interface without reprogramming the node. The application layer includes drivers for each sensor type (I2C, SDI-12, 1-Wire, SPI). Thus, updating the sensor type loads the corresponding driver dynamically.

\begin{figure}[t]
\centering
    \subfloat[]{%
    \includegraphics[width=2.8in]{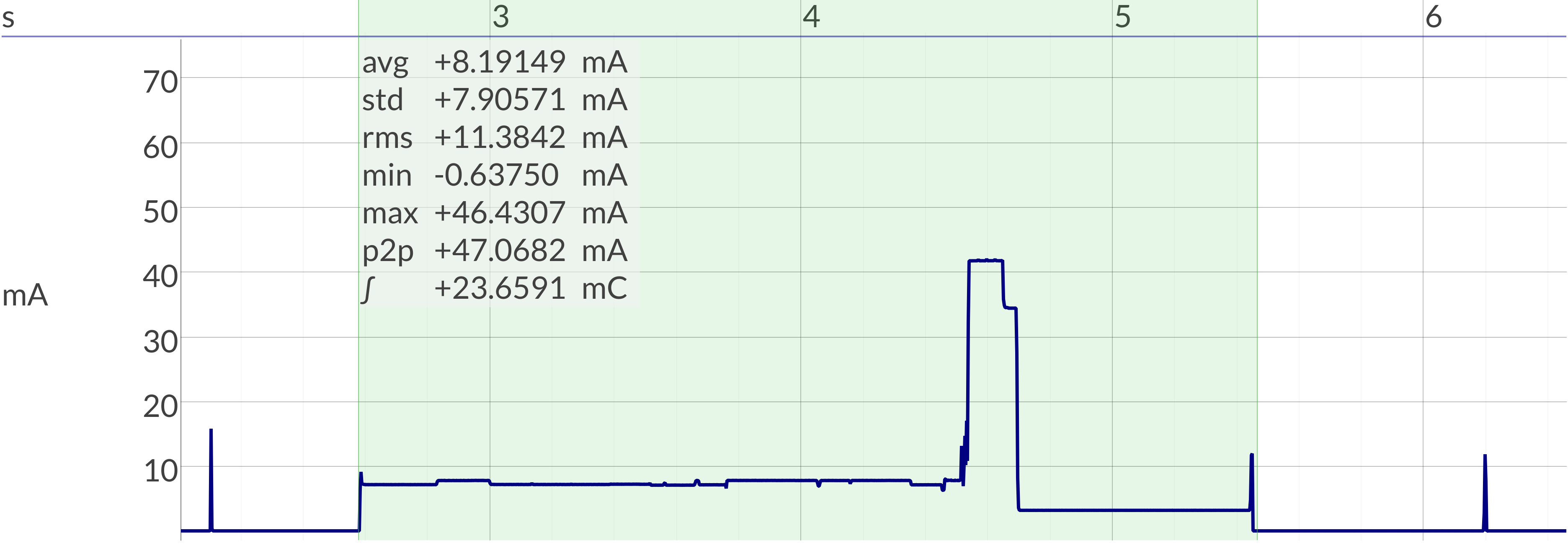}
    \label{fig_power_profile_tx}}
    \quad
    \hspace{-2em}
    \subfloat[]{%
    \includegraphics[width=2.8in]{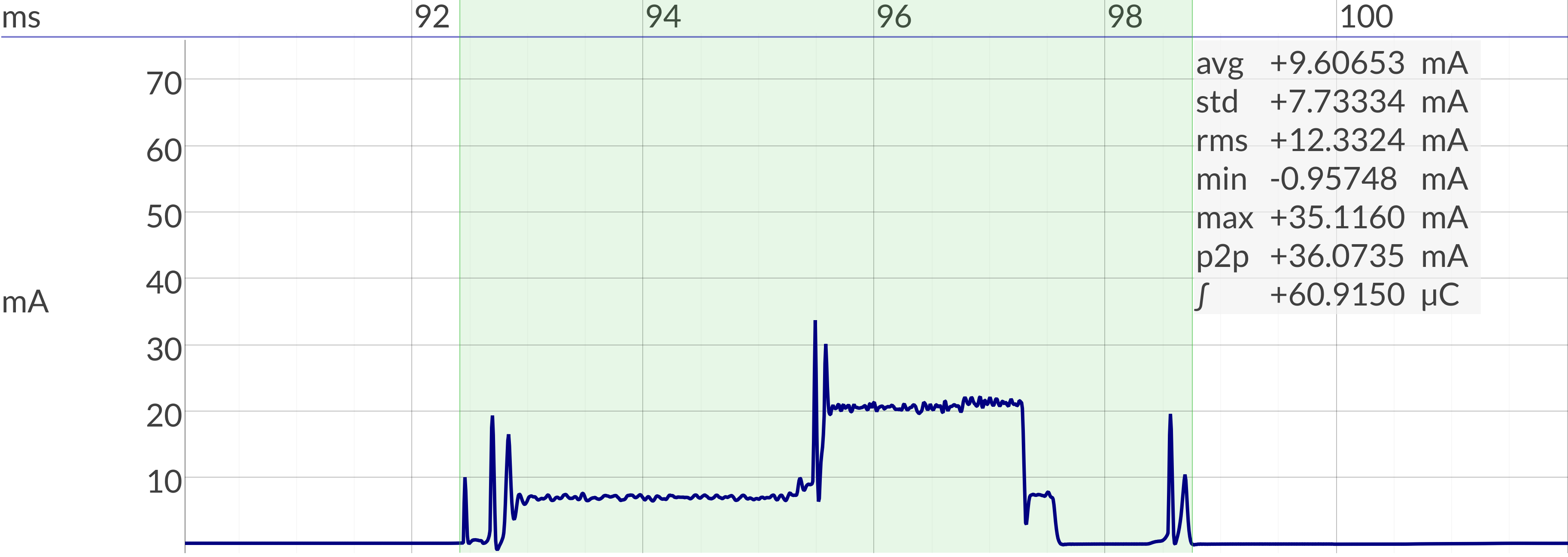}
    \label{fig_power_profile_zoom}}

    \caption{Current consumption of the sensor node at constant \SI{3.3}{\volt} with a temperature sensor connected \protect\subref{fig_power_profile_tx} wake up from sleep, data acquisition and transmission, \protect\subref{fig_power_profile_zoom} low power receive window.}
    \label{fig_sn_power_profile}
\end{figure}

The power consumption of the sensor node after wake-up to acquire and transmit data is shown in Fig.~\ref{fig_power_profile_tx}. It takes around \SI{85.16}{\milli\joule} of energy for the entire   process of data acquisition, processing, and transmission. The sensor nodes sample and transmit sensor data every \num{15}~minutes. Although the application requires only one packet every hour,  oversampling was done to deal with packet losses.  The intermittent low-power wake-ups and sniffing for incoming packets happen \SI{750}{\milli\second} and consume around \SI{201.13}{\micro\joule}.  The spikes on the left and right sides of the highlighted area in Fig.~\ref{fig_power_profile_tx}   indicate these wake-ups, with a zoomed-in version provided in Fig.~\ref{fig_power_profile_zoom}.  Between these energy-consuming tasks, the device remains in sleep mode, consuming approximately \SI{100}{\micro\ampere}.

Using the measurements shown in Fig.~\ref{fig_sn_power_profile}, the total energy consumption of the node per day is around \SI{69.48}{joule}, with the sleep mode, intermittent wake-ups, and the data acquisitions, and transmission, consuming around \SI{43.13}{\joule}, \SI{18.17}{\joule}, and \SI{8.17}{\joule} respectively. Therefore, the lifetime of the device should be around \num{9.70}~years, assuming the entire capacity of the \SI{19}{\ampere\hour} battery is available. However, this is an idealized calculation, as the battery is exposed to extremely low temperatures, which can significantly reduce the available energy.  In Section~\ref{section:discussions}, we discuss the actual lifetime achieved for the nodes.

\subsection{Gateway}
A wireless gateway requires two network interfaces, one for communicating with the low-power sensor nodes and the other for sending data to the backend over an internet connection. As a result, the gateway's power consumption is significantly higher than the sensor nodes, making the choice of network interface and power supply critical.  Since there was NB-IoT connectivity available, we had to rely on  GPRS   for internet access. However, GPRS connection leads to a very high power consumption which cannot be supplied with primary cells. Therefore, we installed a solar power station with \SI{100}{\watt} solar panel and a \SI{100}{\ampere\hour} battery with self-heating to withstand freezing temperatures.  The gateway has an average current consumption of \SI{21.5}{\milli\ampere} at \SI{12}{\volt}, giving a backup time of more than five months in ideal conditions.  

A picture of the deployed gateway is shown in Fig.~\ref{fig_gateway_in_winter}. The gateway uses an OCTA board as the \gls{d7a} module and a Quectel BG~96 module as the GPRS modem.  Similar to the sensor node, the gateway runs the SubIoT stack configured as a \gls{d7a} gateway. However, the gateway does not implement the application layer; instead, it is offloaded to the backend to simplify software complexities.

\subsection{Network Architecture}
\begin{figure}[t]
    \centering
    \includegraphics[width=2.8in]{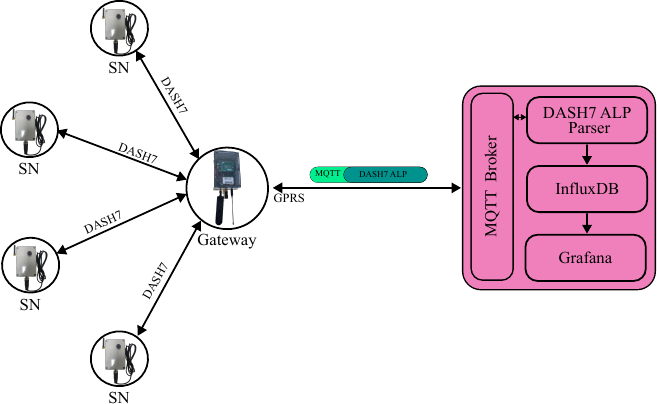}
    \caption{A diagrammatic representation of the \gls{lpwan} and the backend framework for data storage and visualization.}
    \label{fig:network_architecture}
    \vspace{-1em}
\end{figure}

The deployment sites are scattered over a wide area, with each site separated by hilly terrain. Moreover, there is no direct access to power or internet connectivity. Therefore, we decided to deploy an independent star network at each location as shown in Fig.~\ref{fig:network_architecture}.  The gateway connects to the backend which runs the database and visualization framework~(Fig.~\ref{fig:network_architecture}). MQTT facilitates data transfer between the gateway and the backend, InfluxDB stores the collected data, and Grafana serves as the front end.

As stated earlier, the \gls{d7a} gateway does not parse the application layer packets. Instead, it forwards the raw packets to the backend.  The \gls{alp}  of the gateway stack is implemented at the backend in Python using the pyd7a package available from SubIoT.  pyd7a allows splitting the \gls{d7a} stack between a low-end and a high-end device; for instance, a low-power microcontroller and a remote virtual machine.  This brings another advantage that the backend can talk to the remote sensor nodes in a similar way as the gateway would do; i.e., using file names and node IDs.  Therefore, a sensor node in the field can be seamlessly accessed from the backend.  The communication between the application layer and the network layer happens over MQTT, wrapping \gls{d7a} packets in MQTT. 

\section{Deployment} \label{section:deployment}
The deployment was planned to be carried out in three different phases, with each phase covering \gls{gn}~1-3, \gls{gn}~4-5 and \gls{go}, respectively. The first deployment at \gls{gn}~1-3, initially planned for June 2020, was delayed to April 2021 due to COVID-19 and travel restrictions. It covered the first 3 plots containing 18 transects, with 18 soil temperature sensors, one water content sensor, and a gateway. The distance between the gateway and the last node on this site is around \SI{300}{\meter}. After the successful deployment at \gls{gn}~1-3, we tried to combine both the \gls{gn}~4-5 and \gls{go} deployment in another attempt in November 2021. However, the attempt was unsuccessful due to failed gateway installations in adverse weather conditions, although most of the sensor nodes had already been deployed and left untouched. Later, the gateway and solar power station deployment were completed in Aug 2022.  After the successful third phase, we had a total \num{58} sensor nodes and \num{3} gateways active in the field. The gateways in both the \gls{gn}~4-5 and \gls{go} cover nodes in a diameter of approximately \SI{150}{\meter}. In summer 2023, all the sensor nodes from \gls{go} and some from the other two sites were moved to a new experiment site close to \gls{gn}~4-5 and they communicate with the \gls{gn}~4-5  gateway.

\section{ Results and Discussions} \label{section:discussions}
\begin{figure*}[h!]
\centering
    \subfloat[]{%
    \includegraphics[width=2.25in]{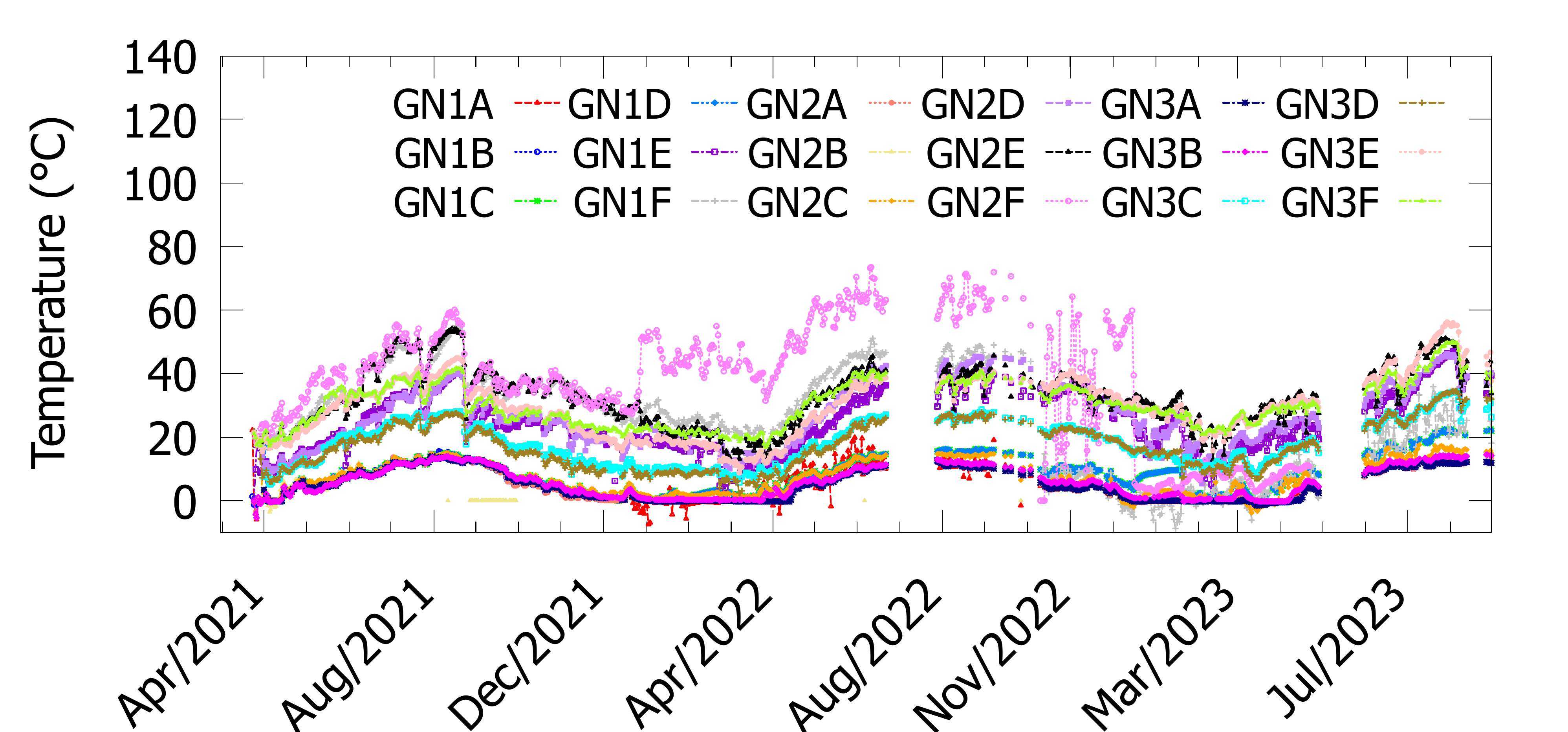}
    \label{fig_gn13_soil_temp_daily}}
    \quad
    \hspace{-1em}
    \subfloat[]{%
    \includegraphics[width=2.25in]{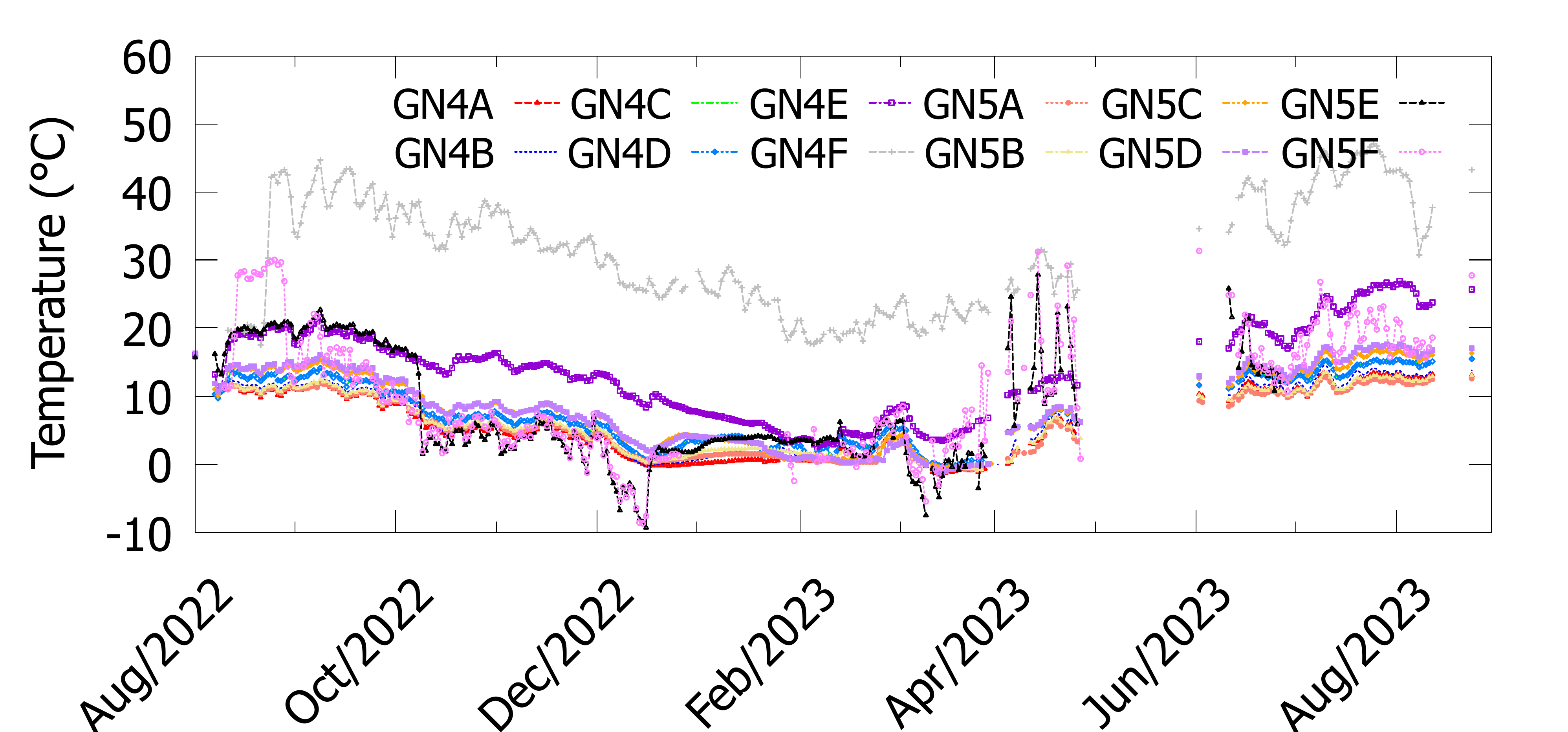}
    \label{fig_gn45_soil_temp_daily}}%
    \subfloat[]{
     \includegraphics[width=1.3in]{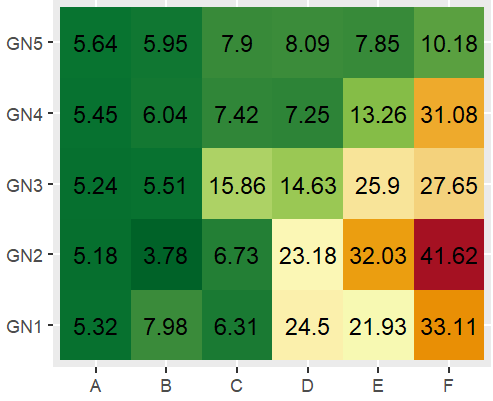}
     \label{fig_gn_soil_temp_annual}}

    \caption{Daily averages of the soil temperature from sensor nodes in \protect\subref{fig_gn13_soil_temp_daily} \gls{gn}~1-3, \protect\subref{fig_gn45_soil_temp_daily} \gls{gn}~4-5, and \protect\subref{fig_gn_soil_temp_annual} heat map of the annual average soil temperature for all the nodes in \gls{gn}.}
    \label{fig_soil_temp}
\end{figure*}

\begin{figure*}[t]
\centering
    \subfloat[]{%
    \includegraphics[width=2.25in]{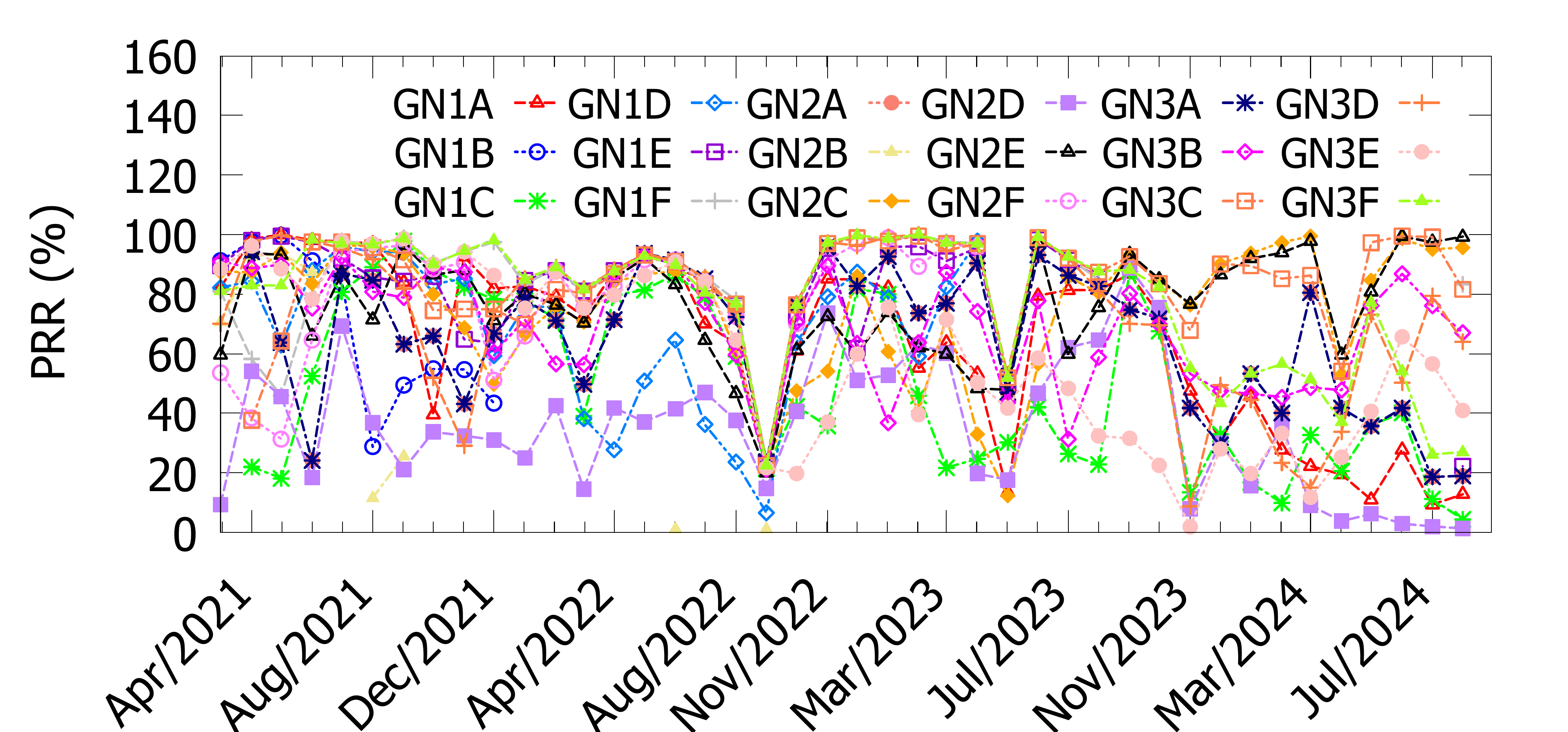}
    \label{fig_gn13_prr_monthly}}
    \quad
    \hspace{-1em}
    \subfloat[]{%
    \includegraphics[width=2.25in]{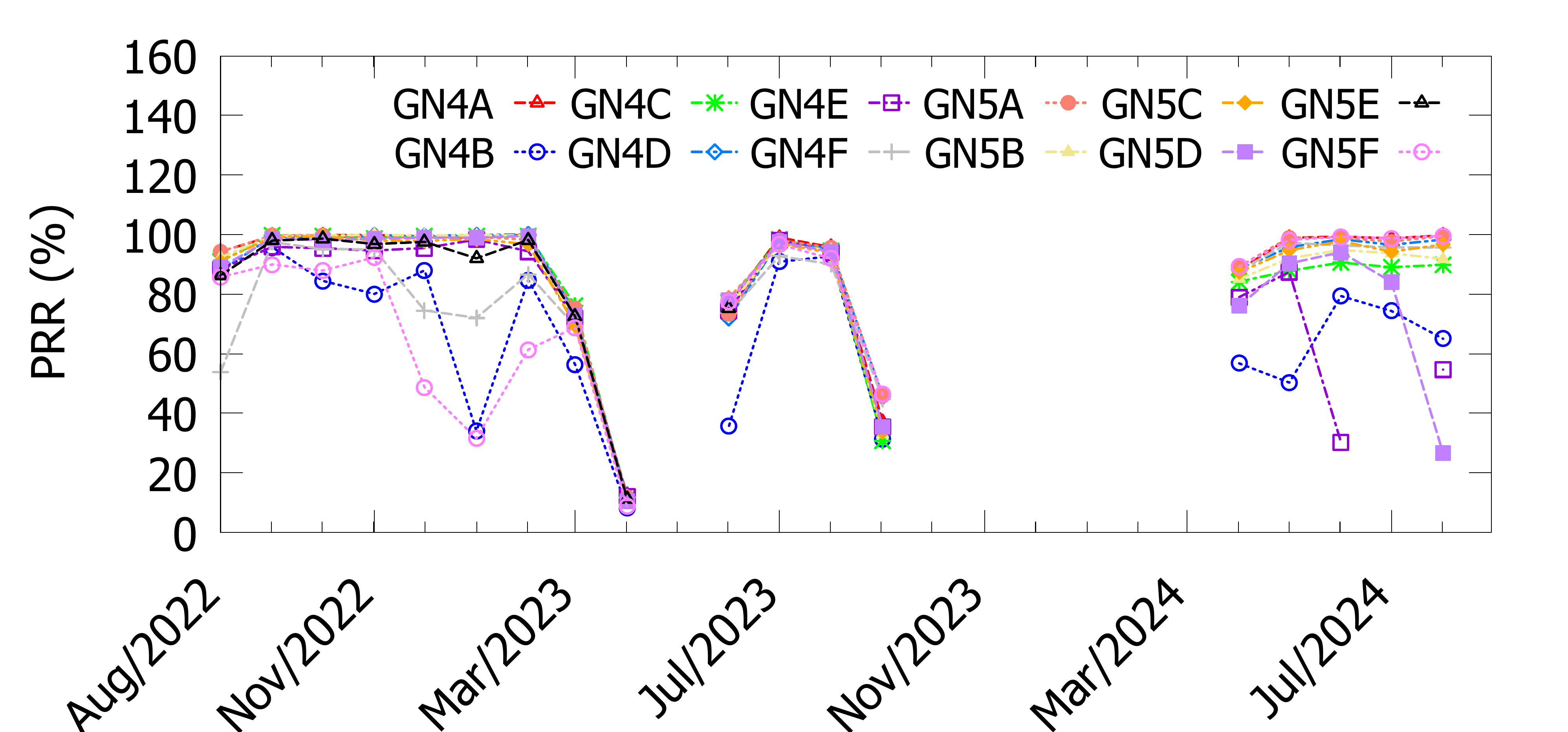}
    \label{fig_gn45_prr_monthly}}%
    \subfloat[]{
     \includegraphics[width=1.2in]{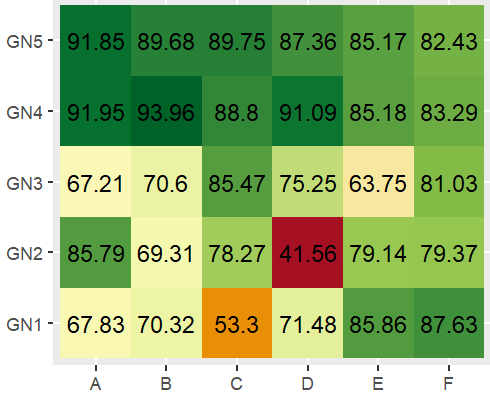}
     \label{fig_prr_overall}}
       \caption{Monthly averages of the \acrshort{prr}~(\%) of the  sensor nodes in \protect\subref{fig_gn13_prr_monthly},  \gls{gn}~1-3 \protect\subref{fig_gn45_prr_monthly} \gls{gn}~4-5, and \protect\subref{fig_prr_overall} heat map of the overall \acrshort{prr} for all the nodes in \gls{gn}.}
    \label{fig_prr}%
\end{figure*}

A decade back, \gls{wsn}  and deployment used to be a challenging task. Ensuring long-term operation and establishing low-power connectivity without depleting batteries created troubles for engineers. However, with the advancements in low-power electronics and connectivity, it has become relatively easy to deploy and maintain \glspl{wsn}. Nevertheless, there is still room for improvement. In this section, we present the key outcomes of the deployment, which has spanned more than three years, and the experiences gained from it.

\textit{Data Collection:} The daily averages of the soil temperature data are plotted in Fig.~\ref{fig_soil_temp}. Please note that we plotted the data until August 2023, since some nodes were relocated for a different experiment. The annual average temperature for each transect in the entire \gls{gn} is depicted in Fig.~\ref{fig_gn_soil_temp_annual}. The warming levels of the soil are visible from the figure. However,   it is worth mentioning that the calculations are based on raw data. We are currently validating the data sets and evaluating the measurements against a set of previously deployed temperature sensors in the field. All the data (soil temperature, air temperature, humidity, weather and RSSI) including the measurements from the weather station and \gls{go} site are available publicly\footnote {https://www.futurearctic.be/data-access}.

\textit{Reliability}: Multiple parameters can be considered to assess the reliability of a network.  Autonomous functioning and the downtime of the network can be primary reliability indicators. As of May 2024, the network at \gls{gn}~1-3 has reached three years in operation. Similarly,  by August 2024, the \gls{gn}~4-5 has reached two years of operation.  We do not consider the \gls{go} site since it was dismantled after a short duration. Over the three years of operation at \gls{gn}~1-3, we experienced minimal network downtime that required a field visit. However, there had been multiple short-duration outages lasting a couple of days to weeks, mostly due to the malfunctioning of the gateways and power or connectivity issues.  Whereas at the \gls{gn}~4-5, we had two major outages which lasted one month or more. In April-May 2023, the network went down for a month and we received no data. As a result, a field visit was conducted, during which it was discovered that the cellular antennas of the gateways had somehow become detached. Though we are unsure how this happened, we presume this was due to the strong winds hitting the antenna flaps.  In another instant, the network became dysfunctional for more than \num{5} months, starting from November 2023. This time, we had a surprising issue as the battery of the solar power station was damaged. Since it was during the winter,  moving heavy batteries to the deployment site was nearly impossible, and we had to wait till the snow subsided. Consequently, the network was rebooted only in April 2024.  A similar issue was reported at the \gls{go} site during the winter of 2023, which further reduced the actual uptime of the \gls{go} network. We encountered the antenna issues at the \gls{go} site as well in November 2022. Following this, the network was not rebooted; instead, all the nodes were relocated.

Although network uptime demonstrates the overall reliability of the network, the \gls{prr} gives a better idea of the node-level reliability.   We estimated the \gls{prr} of each sensor node as the percentage of the total received packets per day to the expected number of packers, i.e., \num{96} packets per day. The monthly averages of the \gls{prr} for each node for the entire deployment duration are plotted in Fig.~\ref{fig_prr}. The previously discussed major and minor network outages can be observed in the graph as disconnected points. From Fig.~\ref{fig_gn45_prr_monthly}, almost all the nodes in \gls{gn}~4-5 have a high \gls{prr} throughout the time, except the months close to the outages. This high \gls{prr}  can be attributed to the relatively low node density and the proximity of nodes to the gateway. The dips in \gls{prr} just before the outage could be the early symptoms of the outage which we missed noticing. In addition, the winter months also recorded a relatively low \gls{prr} for the nodes far from the gateway. 

The nodes in \gls{gn}~1-3 have a rather inferior \gls{prr} compared to \gls{gn}~4-5. We attribute multiple reasons for this result. Firstly,  the gateway and the nodes are separated by a greater distance- more than \SI{300}{\meter}. Additionally, there is a blockage created by spruce trees between the gateway and the site, which is significant during the growing season, i.e., summer. Consequently, a decrease in \gls{prr} can be observed for most of the nodes in the summer months. In addition, the winter months result in another period of reduced \gls{prr} due to snow cover. This site has the highest snow accumulation, causing the nodes to be buried under the snow in the winter.  It can also be observed that a few nodes in \gls{gn}~1-3   have a subpar  \gls{prr}~(Fig.~\ref{fig_prr_overall}). However, since we over-sampled the sensors, we still received enough data to estimate the temperature accurately.

\begin{figure}[h]
\centering
    \subfloat[]{%
    \includegraphics[width=1.65in]{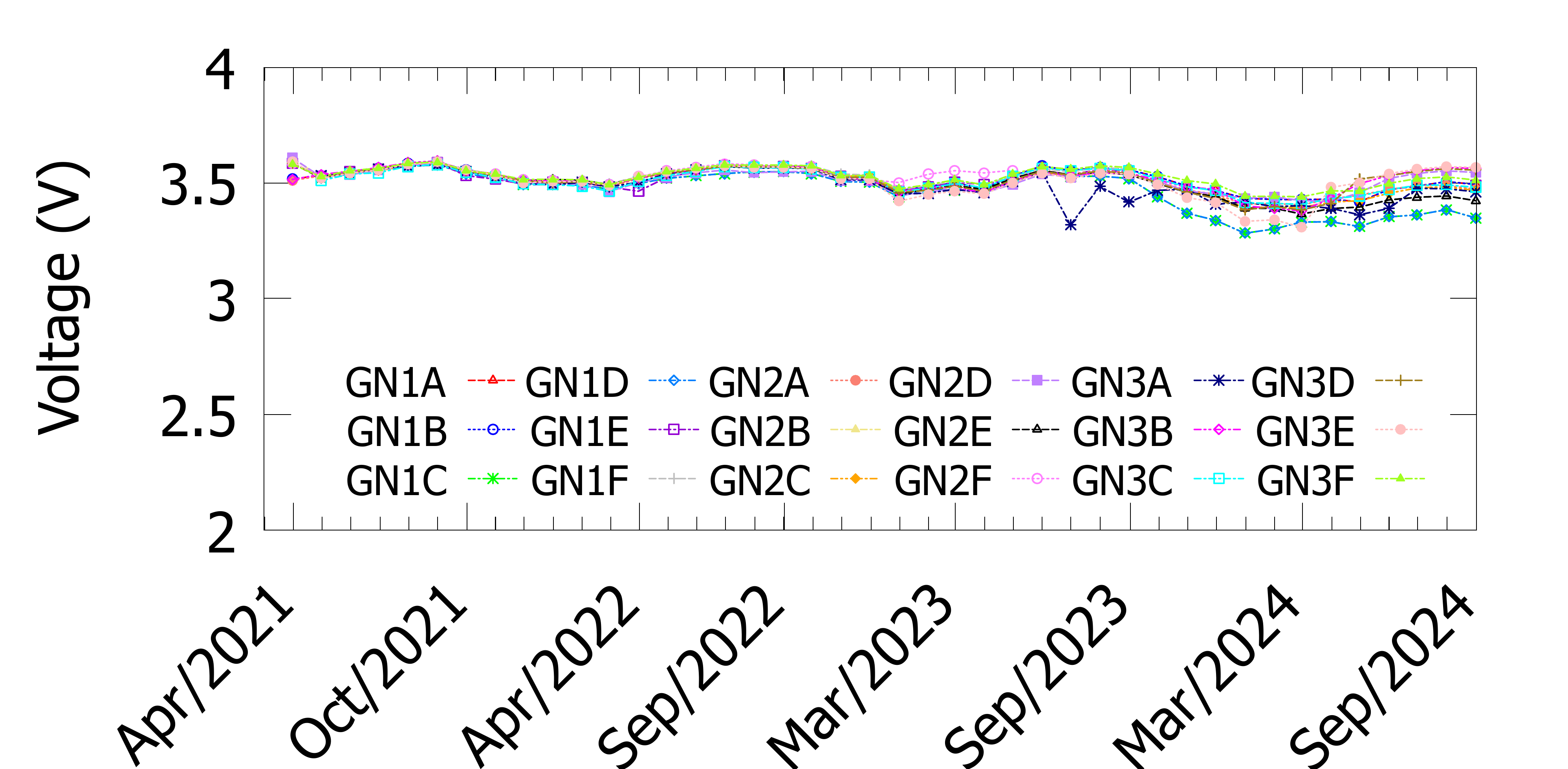}
    \label{fig_gn13_bat_voltage}}
    \hspace{-1.5em} %
    \quad
    \subfloat[]{%
    \includegraphics[width=1.65in]{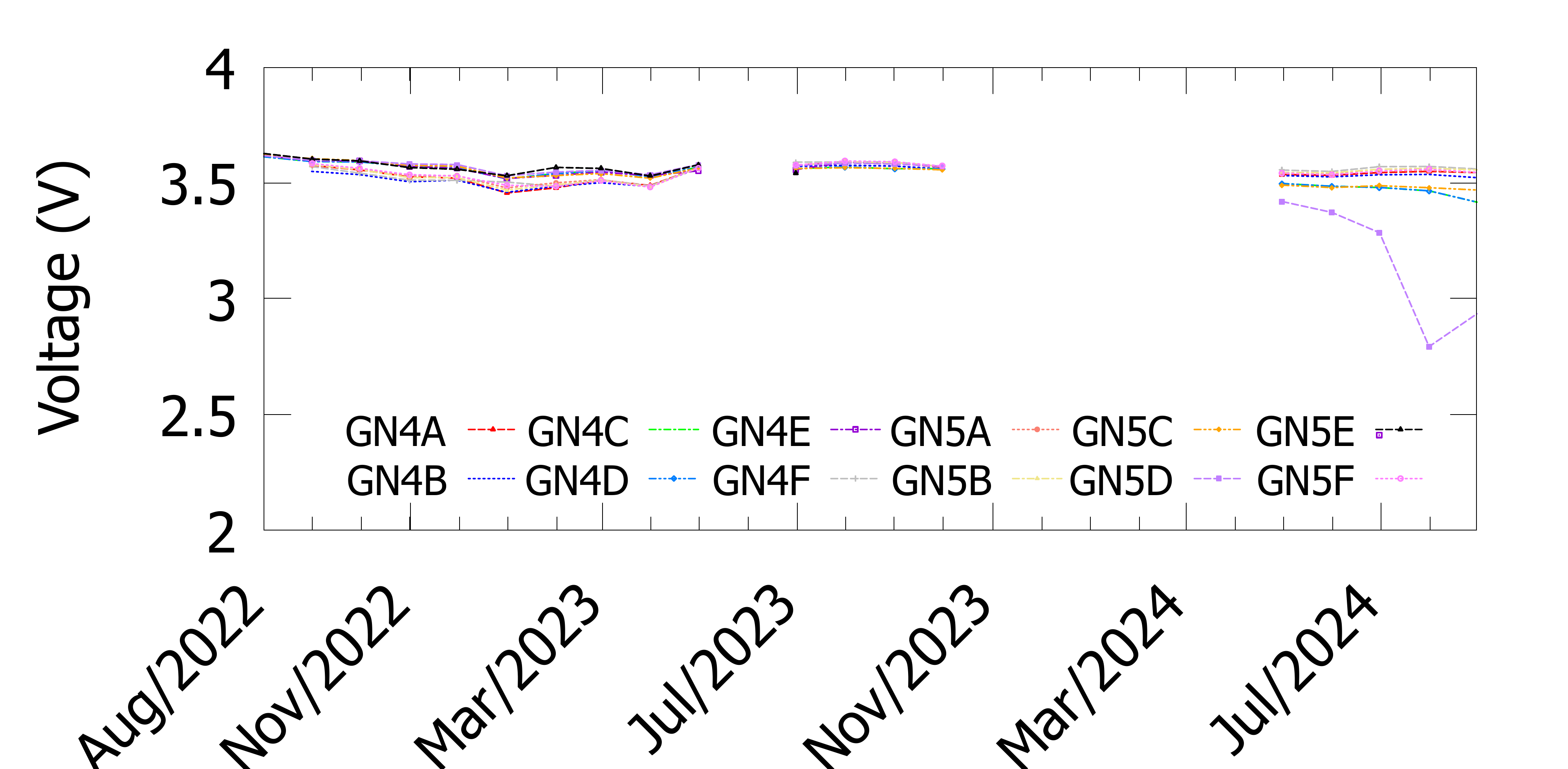}
    \label{fig_gn45_bat_voltage}}
    \caption{Monthly averages of the battery voltage measured for all the sensor nodes in the \protect\subref{fig_gn13_bat_voltage} \gls{gn}~1-3 and \protect\subref{fig_gn45_bat_voltage} \gls{gn}~4-5.}
    \label{fig_sn_bat_voltage}
\end{figure}

\textit{Battery lifetime:} Our initial calculations, based on the power profile and battery capacity, estimated close to  \num{10} years of lifetime for the sensor nodes. However, we observed that this number deviates greatly in real life due to many factors. We have been monitoring the battery voltage of each node, the monthly averages of which are plotted in Fig.~\ref{fig_sn_bat_voltage}. The nodes in \gls{gn}~4-5 still have stable battery voltage, except for a couple of nodes. Whereas the nodes in \gls{gn}~1-3 have already started showing notable fluctuations in battery voltage.  After November 2023, some of the nodes have stopped sending data. We also replaced old batteries on some devices while re-organizing the network in the summer of 2023. An increase in the voltage of some of the nodes (e.g. \gls{gn}2A) is due to the battery replacement. While we are unsure how long the other devices in both \gls{gn} sites would run without a battery replacement, it is certain that a \num{10} year lifetime is nearly impossible. On inspecting the data sheet of the battery~\cite{liscol2_cell}, the highest capacity is achieved when discharged at around \SI{20}{\milli\ampere} and \SI{20}{\celsius}. A decrease in discharge current, as well as operating temperature,  significantly diminishes the capacity. When discharged at \SI{2}{\milli\ampere} and \SI{20}{\celsius}, the achievable capacity is around \SI{11}{\ampere\hour}. Considering that the average annual temperature in Iceland can be less than \SI{10}{\celsius}, a further degradation in capacity may be expected. As of June 2024, we have a total of \num{49} sensor nodes active, with the battery replaced in a few nodes already.

\textit{Feasibility of Energy Harvesting:} We were well aware that energy harvesting could have been a potential solution to the battery lifetime issue. However, we were hesitant to depend on it for several reasons: first, there were no off-the-shelf, proven, and reliable energy harvesting solutions available on the market. Though we could have experimented with energy harvesters, we deliberately avoided this, considering the immediate requirement for data. The availability of high energy density $LiSOCl_{2}$ primary cells in small form factor made the choice easy. Another factor favoring the battery is the lack of reliable ambient energy sources. Solar energy is the preferred option for outdoor deployments; however, the harsh Icelandic winter weather makes it largely inefficient. Nonetheless, over the course of the project, we have uncovered a unique source of energy for outdoor deployments, which is the temperature difference between the soil and the air. While this energy is particularly available at ForHot due to the geothermal activity, it is present even in places without geothermal heating~\cite{soteg}.  One notable feature that sets this energy source apart from solar is its near 24/7 availability, providing power both day and night. However, its power density is often limited, making a multi-source energy harvester that combines both solar and soil-air thermal energy the optimal choice for outdoor deployment in harsh conditions. We have advanced the research in this context and have developed prototypes and proof-of-concept devices that work with soil-air thermal energy~\cite{soteg} and multi-source energy harvesting architectures~\cite{ien}.


\section{Conclusion}
\label{section:conclusion}
We have discussed the design, deployment and maintenance of low-power wireless sensor networks for environment monitoring and sensing at three remote sites in Iceland. The networks, functional since 2021, have achieved a lifespan of over three years with sensor nodes powered by $LiSOCl_{2}$ batteries and remain functional. The mid-range communication protocol \gls{d7a} is proven to be a reliable choice, particularly due to its low-power and efficient downlink capabilities. 


\section*{Acknowledgement}
The authors thank Dr. Páll Sigur$\eth$sson, Prof. Bjarni Di$\eth$rik Sigur$\eth$sson, other FutureArctic colleagues, and the staff at the Agricultural University of Iceland, Hverager$\eth$i for their enormous support and assistance during the deployment and maintenance of the network. The authors also thank the anonymous reviewers of IEEE WCNC’25 and EWSN'24 for their constructive feedback and comments on our work.

This research has received funding from the European Union’s Horizon 2020 research and innovation program under the Marie Sokolowski-Curie grant agreement No 813114.
\bibliographystyle{IEEEtran}
\bibliography{references.bib}
\end{document}